\documentclass[letterpaper,12pt]{article}
\usepackage{amsfonts,latexsym,amssymb,amsmath,algorithmic,epsfig}
\newlength{\jmr}
\newlength{\bjorn}
\newtheorem{q}{Question}
\newtheorem{thm}{Theorem}
\newtheorem{lemma}{Lemma}
\newtheorem{cor}{Corollary}

\renewcommand{\mod}{\mathrm{mod}}
\newtheorem{fph}{Flat Primes Hypothesis (FPH)}

\newtheorem{hensel}{Hensel's Lemma}

\newtheorem{rems}{Remarks} 
\newtheorem{dfn}{Definition}

\newcommand{\sat}{\mathbf{3CNFSAT}}

\newcommand{\thth}{^{\text{\underline{th}}}}

\newcommand{\ord}{{\mathrm{ord}}}

\newcommand{\np}{{\mathbf{NP}}}

\newcommand{\conp}{{\mathbf{coNP}}}

\newcommand{\feas}{{\mathbf{FEAS}}}
\newcommand{\ffp}{{\feas_{\F_\mathrm{primes}}}}
\newcommand{\fqp}{{\feas_{\Q_\mathrm{primes}}}}
\newcommand{\fqpi}{{\feas_{\Q_\mathrm{prime\_ideals}}}}

\newcommand{\bpp}{{\mathbf{BPP}}}
\newcommand{\bqp}{{\mathbf{BQP}}}

\newcommand{\pp}{\mathbf{P}}

\newcommand{\pspa}{\mathbf{PSPACE}}
\newcommand{\expt}{\mathbf{EXPTIME}}
\newcommand{\eps}{\varepsilon}

\newcommand{\F}{\mathbb{F}}
\newcommand{\Q}{\mathbb{Q}}
\newcommand{\R}{\mathbb{R}}
\newcommand{\C}{\mathbb{C}}
\newcommand{\N}{\mathbb{N}}
\newcommand{\Z}{\mathbb{Z}}

\newcommand{\fii}{\varphi}

\newcommand{\ca}{\mathfrak{a}}

\newcommand{\Qn}{\Q^n}

\newcommand{\cC}{{\mathcal{C}}}
\newcommand{\cF}{{\mathcal{F}}}

\newcommand{\cN}{{\mathcal{N}}}

\newcommand{\cP}{{\mathcal{P}}}
\newcommand{\cS}{{\mathcal{S}}}
\newcommand{\cU}{{\mathcal{U}}}

\newcommand{\qed}{$\blacksquare$}
\newcommand{\dia}{$\diamond$}
\newcommand{\size}{\mathrm{size}}

\newlength{\hwl}
\begin{document}
\title{\mbox{}\\
\vspace{-1in} A Number Theoretic Interpolation Between Quantum and 
Classical Complexity Classes} 
\author{
J. Maurice Rojas\thanks{Department of Mathematics,
Texas A\&M University, TAMU 3368,
College Station, Texas \ 77843-3368,
USA.  {\tt rojas@math.tamu.edu } , {\tt
www.math.tamu.edu/\~{}rojas} \ .
Partially supported by NSF individual grant DMS-0211458, NSF CAREER grant 
DMS-0349309, and Sandia National Laboratories.} 
} 

\date{\today} 

\maketitle

\vspace{-.7cm}

\begin{abstract}  
We reveal a natural algebraic problem whose complexity appears to interpolate 
between the well-known complexity classes $\bqp$ and $\np$:\\
$\star$ Decide whether a univariate polynomial with exactly $m$ 
monomial terms has a $p$-adic\\
\mbox{}\hspace{.3cm}rational root.\\ 
In particular, we show that while ($\star$) is doable 
in quantum randomized polynomial time when $m\!=\!2$ (and 
no classical randomized polynomial time algorithm is known), ($\star$) is 
nearly $\np$-hard for general $m$: Under a 
plausible hypothesis involving primes in arithmetic progression 
(implied by the Generalized Riemann Hypothesis for certain cyclotomic fields), 
a randomized polynomial time algorithm for ($\star$) 
would imply the widely disbelieved inclusion 
$\np\!\subseteq\!\bpp$. This type of quantum/classical  
interpolation phenomenon appears to new. 
\end{abstract} 

\section{Introduction and Main Results} 
Thanks to quantum computation, we now have exponential 
speed-ups for important practical problems such as Integer Factoring and 
Discrete Logarithm \cite{shor}. However, a fundamental open question 
that remains is whether there are any {\bf $\np$-complete} problems admitting 
exponential speed-ups via quantum computation. Succinctly, this is 
the $\np\!\stackrel{?}{\subseteq}\!\bqp$ question, and a positive 
answer would imply that quantum computation can provide  
efficient algorithms for a myriad of problems that have occupied 
practicioners in optimization and computer science for decades \cite{bv}. 
(The classic reference \cite{gj} lists dozens of such problems, and 
we briefly review the aforementioned complexity classes in Section 2 below.) 
However, the truth of the inclusion $\np\!\subseteq\!\bqp$ is currently 
unknown and widely disbelieved (as of early 2006).  

We present an algebraic approach to this question by illustrating a 
problem, involving sparse polynomials over $\Q_p$ (the $p$-adic rationals), 
whose complexity appears to interpolate between the complexity classes 
$\bqp$ and $\np$. Our results thus suggest that sparse polynomials can shed 
light on the difference between $\bqp$ and $\np$. Indeed, one consequence 
of our results is a new family of problems which admit (or are likely  
to admit) $\bqp$ algorithms. Also, in addition to providing a new 
complexity limit for factoring polynomials over $\Q_p$, we can address 
questions posed earlier by Cox \cite{cox}, and Karpinski and Shparlinski 
\cite{squarefree}, regarding sparse polynomials over finite fields.

Let us first review some necessary terminology: 
For any ring $R$ containing the integers $\Z$, let $\feas_R$ --- the
{\bf $R$-feasibility problem} --- denote the problem of deciding whether a
given system of polynomials $f_1,\ldots,f_k$ chosen from
$\Z[x_1,\ldots,x_n]$ has a root in $R^n$.  Observe then that
$\feas_\R$ and $\feas_\Q$ are respectively the central problems of 
algorithmic real algebraic geometry and algorithmic arithmetic geometry (see 
Section \ref{sub:padic} below for further details). 

To measure the ``size'' of an input polynomial in our complexity estimates, 
we will essentially count 
just the number of bits needed to write down the coefficients 
and exponents in its monomial term expansion. This is 
the {\bf sparse} input size, as opposed to the ``dense'' input size 
used frequently in computational algebra. 
\begin{dfn}
Let $f(x)\!:=\!\sum^m_{i=1} c_ix^{a_i}\!\in\!\Z[x_1,\ldots,x_n]$
where $x^{a_i}\!:=\!x^{a_{1i}}_1\cdots x^{a_{ni}}_n$, $c_i\!\neq\!0$ for all
$i$, and the $a_i$ are distinct. We call such an $f$ an
{\bf $\pmb{n}$-variate $\pmb{m}$-nomial}. 
Also let\\
\mbox{}\hfill $\size(f)\!:=\!\sum^m_{i=1}\left(1+
\lceil\log_2(2+|c_i|)\rceil+\lceil\log_2(2+|a_{1,i}|)\rceil+\cdots+\lceil
\log_2(2+|a_{n,i}|)\rceil\right)$,\hfill\mbox{}\\
and $\size_p(f)\!:=\!\size(f)+\log(2+p)$. (We also extend $\size$, and thereby 
$\size_p$, additively to polynomial systems.)
Finally, for any collection $\cF$ of polynomial systems with
integer coefficients, let $\feas_R(\cF)$ denote the natural
restriction of $\feas_R$ to inputs in $\cF$. \dia
\end{dfn} 
Observe that $\size(a+bx^{99}+cx^d)\!=\!\Theta(\log d)$ if we fix $a,b,c$, 
so the degree of a polynomial can sometimes be exponential in its 
sparse size. Since it is not hard to show that $\feas_{\Q_p}(\cU_2)\!\in\!\pp$ 
when $p$ is fixed (cf.\ Section 3 below), it will be more natural to 
take the size of an input prime $p$ into account as well. 
\begin{dfn} Let $\fqp$ (resp.\ $\fqp(\cF)$) denote the
union of problems \linebreak $\bigcup\limits_{p \text{ prime}}\feas_{\Q_p}$
(resp.\ $\bigcup\limits_{p
\text{ prime}}\feas_{\Q_p}(\cF)$), so that a prime $p$ is also part of the 
input, and the underlying input size is $\size_p$.
Also let $Q_n$ denote the product of the first $n$ primes 
and define 
$\cU_m\!:=\!\{f \!\in\!\Z[x_1] \; | \; f \text{ is an $m$-nomial}\}$. 
\dia   
\end{dfn}
Observe that $\Z[x_1]$ is thus the disjoint union 
$\bigsqcup_{m\!\geq\!0}\cU_m$. 
Our results will make use of the following plausible 
number-theoretic hypothesis. 
\begin{fph}
\scalebox{.94}[1]{Following the notation above, there are absolute constants} 
\linebreak  
\scalebox{.92}[1]{$C'\!\geq\!C\!\geq\!1$ such that for any $n\!\in\!\N$, 
the set $\{1+kQ_n\; | \; k\!\in\!\{1, \ldots,2^{n^C}\}\}$ contains at least 
$\frac{2^{n^C}}{n^{C'}}$ primes.} 
\end{fph}
Assumptions at least as strong as FPH are routinely used, and widely
believed, in the cryptology and algorithmic number theory communities 
(see, e.g., \cite{miller,miha,koiran,jcs,hallgren}).
In particular, we will see in Section \ref{sub:riemann} below how FPH is 
implied by the
Generalized Riemann Hypothesis (GRH) for the number fields
$\{\Q(\omega_{Q_n})\}_{n\in\N}$, where $\omega_M$
denotes a primitive $M\thth$ root of unity\footnote{i.e., 
a complex number $\omega_M$ with $\omega^M_M\!=\!1$; and 
$\omega^d_M\!=\!1 \Longrightarrow M|d$}, but can still hold under certain
failures of the latter hypotheses.
\begin{thm}
\label{thm:me}
Following the notation above, $\fqp(\cU_2)\!\in\!\bqp$. However, assuming
the truth of FPH, if $\fqp(\Z[x_1])\!\in\!\cC$ for some complexity class
$\cC$, then $\np\!\subseteq\!\bpp\cup\cC$.
In particular, assuming the truth of FPH, 
$\fqp(\Z[x_1])\!\in\!\bqp \Longrightarrow \np\!\subseteq\!\bqp$.
\end{thm}

\noindent 
Recall that a univariate polynomial has a root in a field $K$ iff it 
possesses a degree $1$ factor with coefficients in $K$. 
Independent of its connection to quantum computing, Theorem 1 
thus provides a new complexity limit for polynomial
factorization over $\Q_p[x_1]$. In particular, Theorem 1 shows that 
finding even just the low degree ($p$-adic) factors for {\bf sparse} 
polynomials (with $p$ varying) is likely {\bf not} doable in randomized 
polynomial time. This complements Chistov's earlier deterministic polynomial 
time algorithm for dense polynomials and fixed $p$ \cite{chistov}. Theorem 1 
also provides an interesting contrast to earlier work of Lenstra 
\cite{lenstra1}, who showed 
that one can at least find all {\bf low} degree factors (in $\Q[x_1]$) of a 
sparse polynomial in polynomial time. 
\begin{rems}
While it has been known since the late 1990's that $\fqp\!\in\!\expt$
\cite{mw1,mw2} (relative to our notion of input size), we are unaware of any
earlier algorithms yielding $\fqp(\cF)\!\in\!\bqp$, for any non-trivial family
of polynomial systems $\cF$. Also, while it is not hard to show that
$\fqp$ is $\np$-hard from scratch, there appear to be no earlier
results indicating the smallest $n$ such that $\fqp(\Z[x_1,\ldots,x_n])$
is $\np$-hard. \dia 
\end{rems}

As for the quantum side of Theorem 1, 
the author is unaware of any other natural algebraic problem that interpolates 
between $\bqp$ and $\np$ in the sense above. 
Moreover, since the exact complexity of the
problems $\{\fqp(\cU_m)\}_{m\geq 3}$ is currently unknown, a $\bqp$ algorithm
for any of these problems would yield a new family of algebraic problems ---
distinct from Integer Factoring or Discrete Logarithm ---
admitting an exponential quantum speed-up over classical methods.

The only other problem known to interpolate between $\bqp$ and  
{\bf some} classical complexity class arises from very recent results on 
the complexity of approximating a certain braid invariant --- the famous
Jones polynomial, for certain classes of braids, evaluated at an $n\thth$ root 
of unity --- and involves a complexity class (apparently) higher than $\np$. 
In brief: (1) seminal work of Freedman, Kitaev, Larsen, and Wang shows 
that such approximations can simulate any $\bqp$ computation, 
already for $n\!=\!5$ \cite{fkw,flw}, 
(2) \cite{dorit} gives a $\bqp$ algorithm that computes an 
additive approximation for arbitrary $n$, 
and (3) \cite{yard} shows that for arbitrary $n$, computing the most 
significant bit of the absolute value of the Jones polynomial is 
$\pp\pp$-hard.  (Recall that $\bqp\cup\np\cup\conp\!\subseteq\!\pp\pp$.)  
Our results thus provide a new alternative source for quantum/classical 
complexity interpolation. 

Let $\ffp$ denote the obvious finite field analogue of $\fqp$. 
While we do not yet know whether $\fqp(\cU_2)$ is 
$\bqp$-complete in any rigourous sense, we point out that 
$\fqp(\cU_2)$ is polynomial-time equivalent to $\ffp(\cU_2)$ 
(cf.\ Section 3 below), and the inclusion $\ffp(\cU_2)\!\stackrel{?}
{\in}\!\bpp$ is a 
well-known, decades-old open problem from algorithmic number theory (see, 
e.g., \cite[Ch.\ 7]{bs} and \cite{gao}). Note also that the 
$\bqp$-completeness of Integer Factoring and Discrete Logarithm are open 
questions as well.  

One can also naturally ask if detecting a {\bf degenerate} root in $\Q_p$ for 
$f$ (i.e., a degree $1$ factor over $\Q_p$ whose square also divides $f$) 
is as hard as detecting arbitary roots in $\Q_p$. 
Via our techniques, we can easily   
prove essentially the same complexity lower-bound as above for the 
latter problem. 
\begin{cor} 
\label{cor:squarefree} 
Using $\size_p(f)$ as our notion of input size, suppose 
we can decide for any input prime $p$ and $f\!\in\!\Z[x_1]$ whether  
$f$ is divisible by the square of a degree $1$ polynomial in $\Q_p[x_1]$, 
within some complexity class $\cC$. Then, assuming the truth of FPH, 
$\np\!\subseteq\!\cC\cup\bpp$. 
\end{cor} 

\noindent 
Let $\F_p$ denote the finite field with $p$ elements. Corollary 1 
then complements an analogous earlier result of Karpinski and Shparlinski 
(independent of the truth of FPH) for detecting degenerate 
roots in $\C$ and the algebraic closure of $\F_p$.
 
Note also that while the truth of GRH usually implies algorithmic 
speed-ups (in contexts such as primality testing \cite{miller}, 
complex dimension computation \cite{koiran}, detection of 
rational points \cite{jcs}, or 
class group computation \cite{hallgren}), Theorem \ref{thm:me} and 
Corollary 1 instead 
reveal complexity {\bf speed-limits} implied by GRH. 

\subsection{Open Questions and the Relevance of Ultrametric Complexity} 
\label{sub:padic} 
Complexity results over one ring sometimes 
inspire and motivate analogous results over other rings.  
An important early instance of such a transfer was the work of 
Paul Cohen, on quantifier elimination over $\R$ and $\Q_p$ \cite{cohen}. 
To close this introduction, let us briefly review how results over $\Q_p$ 
can be useful over $\Q$, and then raise some natural questions arising from 
our main results. 

First, recall that the decidability of $\feas_\Q$ is a major open problem:    
decidability for the special case of cubic polynomials in two variables  
would already be enough to yield significant new results in the direction of 
the Birch-Swinnerton-Dyer conjecture (see, e.g., \cite[Ch.\ 8]{silverman}), 
and the latter conjecture is central in modern number theory (see, e.g., 
\cite{hs}). The fact that $\feas_\Z$ is undecidable is the famous negative 
solution of Hilbert's Tenth Problem, due to Matiyasevitch and Davis, Putnam, 
and Robinson \cite{oldmat,h10}, and is sometimes taken as evidence that 
$\feas_\Q$ may be undecidable as well (see also \cite{bjornlarge}). 

From a more positive direction, much work has gone into using 
$p$-adic methods to find an algorithm for $\feas_\Q(\Z[x,y])$ 
(i.e., deciding the existence of rational points on algebraic curves), via 
extensions of the {\bf Hasse Principle}\footnote{The Hasse Principle is the 
assumption that an equation $F(x_1,\ldots,x_n)\!=\!0$ having roots in 
$\Q^n_p$ for all primes $p$ must have a root in $\Qn$ as well. The Hasse 
Principle is a theorem for quadratic polynomials, is conjectured to hold for 
equations defining smooth plane curves, but fails in subtle ways for cubic 
polynomials (see, e.g., \cite{bjornhasse1}). } 
(see, e.g., \cite{bjornhasse2,bjornbm}). 
Algorithmic results over the $p$-adics are also central in 
many other computational results: 
polynomial time factoring algorithms over $\Q[x_1]$ \cite{lll}, 
computational complexity \cite{antsv}, and elliptic curve cryptography 
\cite{lauder}. 

Our results thus provide another step toward understanding 
the complexity of solving polynomial equations over $\Q_p$, and 
reveal yet another connection between quantum complexity and number theory. 
Let us now consider some possible extensions of our results. 
\begin{q} 
Is $\ffp(\Z[x_1])$ $\np$-hard?
\end{q} 
\begin{q} 
Given a prime $p$ and an $f\!\in\!\F_p[x_1]$, is it $\np$-hard to 
decide whether $f$ is divisible by the square of a degree $1$ polynomial in 
$\F_p[x_1]$ (relative to $\size_p(f)$)? 
\end{q} 

\noindent 
David A.\ Cox asked the author whether $\ffp(\Z[x_1])\!\stackrel{?}{\in}\!\pp$ 
around August 2004 \cite{cox}, 
and Erich Kaltofen posed a variant of Question 1 --- 
$\ffp(\cU_3)\!\stackrel{?}{\in}\!\pp$ 
--- a bit earlier in \cite{kaltofen}. Karpinski and Shparlinski raised 
Question 2 toward the end of \cite{squarefree}. 
Since Hensel's Lemma (cf.\ Section 2 below) allows one to find roots in $\Q_p$ 
via computations in the rings $\Z/p^\ell\Z$, 
Theorem \ref{thm:me} thus provides some evidence toward positive  
answers for Questions 1 and 2.  
Note in particular that a positive answer to Question 1 would 
provide a definitive complexity lower bound for polynomial 
factorization over $\F_p[x_1]$, since randomized polynomial 
time algorithms (relative to the {\bf dense} encoding) are 
already known (e.g., Berlekamp's algorithm \cite[Sec.\ 7.4]{bs}). 

On a more speculative note, one may wonder if quantum computation 
can produce new speed-ups by circumventing the dependence of certain 
algorithms on GRH. This is motivated by Hallgren's recent discovery of 
a $\bqp$ algorithm for deciding whether the class number of a number field 
of constant degree is equal to a given integer \cite{hallgren}: The 
best classical complexity upper bound for the latter problem is 
$\np\cap\conp$, obtainable so far only under the assumption of 
GRH \cite{buchmann,mccurley}. Unfortunately, the precise relation 
between $\bqp$ and $\np\cap\conp$ is not clear. However, could it be that 
quantum computation can expunge the need for GRH in an even more direct way?  
For instance: 
\begin{q} 
Is there a quantum algorithm which generates, 
within a number of qubit operations polynomial in $n$, 
a prime of the form $kQ_n+1$ with probability $>\!\frac{2}{3}$?  
\end{q} 

Indeed, it is natural to try to remove the dependence of 
our main results on the hypothesis FPH. Here is one possible 
route. 
\begin{q} 
Let $\fqpi$ denote the obvious generalization of $\fqp$ to 
arbitrary finite algebraic extensions of the fields $\{\Q_p\}_{p \text{ a 
prime}}$. Then $\fqpi(\Z[x_1])$ is $\np$-hard, independent of 
FPH. 
\end{q} 

\noindent 
We are currently pursuing a solution to the last question. In particular, 
it appears likely that $\fqpi(\cU_2)\!\in\!\bqp$. 

Our main results are proved mostly in Section \ref{sec:proofs1}, after
the development of some necessary theory in Section \ref{sec:proofs} below. 
For the convenience of the reader, we recall the definitions 
of all relevant complexity classes and review certain types of Generalized 
Riemann Hypotheses. 

\section{Background and Ancillary Results} 
\label{sec:proofs}
Recall the containments of complexity classes
$\pp\!\subseteq\!\bpp\!\subseteq\!\bqp\!\subseteq\!\pp\pp\!\subseteq\!\pspa$ 
and $\pp\!\subseteq\!\np\cap\conp\!\subseteq\!\np\cup\conp\!\subseteq\!\pp\pp$, 
and the fact that the properness 
of {\bf every} preceding containment is a major open problem \cite{papa,bv}. 
We briefly review the definitions of the aforementioned complexity classes 
below (see \cite{papa,bv} for a full and rigourous treatment): 
\begin{itemize} 
\item[$\pp$]{ The family of decision problems which can be done within
(classical) polynomial-time. }
\item[$\bpp$]{ The family of decision problems admitting (classical) 
randomized polynomial-time algorithms that terminate with an answer that is 
correct with probability at least\footnote{It is easily shown that we can 
replace $\frac{2}{3}$ by any constant strictly greater than $\frac{1}{2}$
and still obtain the same family of problems \cite{papa}.} $\frac{2}{3}$.}
\item[$\bqp$]{ The family of decision problems admitting {\bf quantum} 
randomized polynomial-time algorithms that terminate with an answer that is 
correct with probability at least$^3$ $\frac{2}{3}$ \cite{bv}. }
\item[$\np$]{ The family of decision problems where a {\tt ``Yes''} answer can
be {\bf certified} within (classical) polynomial-time.}
\item[$\conp$]{ The family of decision problems where a {\tt ``No''} answer 
can be {\bf certified} within (classical) polynomial-time.}
\item[$\pp\pp$]{ The family of decision problems admitting (classical) 
randomized polynomial-time algorithms that terminate with an answer that is 
correct with probability strictly greater than $\frac{1}{2}$.}
\item[$\pspa$]{ The family of decision problems solvable within
polynomial-time, provided a number of processors exponential in
the input size is allowed. }
\end{itemize} 

Now recall that $\sat$ is the famous seminal $\np$-complete problem \cite{gj} 
which consists of deciding whether a Boolean sentence of the form 
$B(X)=C_1(X)\wedge \cdots \wedge C_k(X)$ has a
satisfying assignment, where $C_i$ is of one of the following forms:\\
\mbox{}\hfill $X_i\vee X_j \vee X_k$, \
$\neg X_i\vee X_j \vee X_k$, \
$\neg X_i\vee \neg X_j \vee X_k$,  \
$\neg X_i\vee \neg X_j \vee \neg X_k$, \hfill \mbox{}\\
$i,j,k\!\in\![3n]$, and a satisfying assigment consists of
an assigment of values from $\{0,1\}$ to the variables
$X_1,\ldots,X_{3n}$ which makes the equality $B(X)\!=\!1$ true.\footnote{
Throughout this paper, for Boolean expressions, we will always identify $0$ 
with {\tt ``False''} and  $1$ with {\tt ``True''}.}  
Each $C_i$ is called a {\bf clause}. 

We will need a clever reduction from $\sat$ to 
feasibility testing for univariate polynomial systems over certain fields. 
\begin{dfn}  
\label{dfn:plai} 
Letting $Q_n$ denote the product of the first $n$ primes, let 
us inductively define a homomorphism $\cP_n$ --- 
the {\bf ($n\thth$) Plaisted morphism} --- from certain Boolean 
polynomials in the variables $X_1,\ldots,X_n$ to $\Z[x_1]$, as follows: 
(1) $\cP_n(0)\!:=\!1$, (2) $\cP_n(X_i)\!:=\!x^{Q_n/p_i}_1-1$, 
(3) $\cP_n(\neg B)\!:=\!\frac{x^{Q_n}_1-1}{\cP_n(B)}$, for any 
Boolean polynomial $B$ for which $\cP_n(B)$ has already been defined, 
\linebreak (4) $\cP_n(B_1\vee B_2)\!:=\!\mathrm{lcm}(\cP_n(B_1),\cP_n(B_2))$, 
for any Boolean polynomials $B_1$ and $B_2$ for which $\cP_n(B_1)$ and 
$\cP_n(B_2)$ have already been defined. \dia 
\end{dfn} 
\begin{lemma} 
\label{lemma:plai} 
For all $n\!\in\!\N$ and all clauses $C(X_i,X_j,X_k)$ with $i,j,k\!\leq\!n$, 
we have $\size(\cP_n(C))\!=\!O(n^2)$. Furthermore, if $K$ is any field 
possessing $Q_n$ distinct ${Q_n}\thth$ roots of unity, then 
a $\sat$ instance $B(X)\!:=\!C_1(X)\wedge \cdots \wedge C_k(X)$ has a 
satisfying assignment iff the zero set in $K$ of the polynomial system 
$F_B\!:=\!(\cP_n(C_1), \ldots,\cP_n(C_k))$ has a root $\zeta$ satisfying 
$\zeta^{Q_n}-1$. \qed 
\end{lemma} 

\noindent
David Alan Plaisted proved the special case $K\!=\!\C$ of the 
above lemma in \cite{plaisted}. His proof extends with no difficulty 
whatsoever to the more general family of fields detailed above. Other 
than a slightly earlier (and independent) observation of Kaltofen 
and Koiran \cite{kk}, we are unaware of any other variant of Plaisted's 
reduction involving a field other than $\C$. 

Let us recall a version of Hensel's Lemma sufficiently 
general for our proof of Theorem \ref{thm:me}, along 
with a useful characterization of certain finite rings. 
Recall that for any ring $R$, $R^*$ is the group of multiplicatively 
invertible elements of $R$. 
\begin{hensel} 
(See, e.g., \cite[Pg.\ 48]{robert}.)   
Suppose $f\!\in\!\Z_p[x_1]$ and $x\!\in\!\Z_p$ 
satisfies $f(x)\!\equiv\!0 \ (\mod \ p^\ell)$ and 
$\ord_p f'(x)\!<\!\frac{\ell}{2}$. Then there is  
a root $\zeta\!\in\!\Z_p$ of $f$ with $\zeta\!\equiv\!x \ 
(\mod \ p^{\ell-\ord_p f'(x)})$ and $\ord_p f'(\zeta)\!=\!\ord_pf'(x)$. 
\qed 
\end{hensel}
\begin{lemma}
\label{lemma:fp}
Given any cyclic group $G$, $a\!\in\!G$, and an integer $d$, the equation 
$x^d\!=\!a$ has a solution iff the order of $a$ divides 
$\frac{\#G}{\gcd(d,\#G)}$. In particular, $F^*_q$ is cyclic for 
any prime power $q$, and $(\Z/p^\ell\Z)^*$ is cyclic 
for any $(p,\ell)$ with $p$ an odd prime or $\ell\!\leq\!2$. 
Finally, for $\ell\!\geq\!3$,\linebreak $(\Z/2^\ell\Z)^*\!\cong\!
\{-1,1\}\times\{1,5,5^2,5^3,\ldots,5^{2^{\ell-2}-1} \ \mod \ 2^\ell\}$. 
\qed
\end{lemma}

\noindent
The last lemma is standard (see, e.g., \cite[Ch.\ 5]{bs}). 

We will also need the following result on an efficient randomized reduction  
of $\feas_K(\Z[x_1]^k)$ to $\feas_K(\Z[x_1]^2)$. Recall that 
$\C_p$ --- the $p$-adic complex numbers --- is the metric closure of the 
algebraic closure of $\Q_p$, and $\C_p$ is algebraically closed. 
\begin{lemma}
\label{lemma:gh}
Suppose $f_1,\ldots,f_k\!\in\!\Z[x_1]\setminus\{0\}$ are
polynomials of degree $\leq\!d$, with $k\!\geq\!3$.
Also let $Z_K(f_1,\ldots,f_k)$ denote the
set of common zeroes of $f_1,\ldots,f_k$ in some field $K$. Then, if
$a\!=\!(a_1,\ldots,a_k)$ and $b\!=\!(b_1,\ldots,b_k)$ are chosen uniformly
randomly from $\{1,\ldots,18dk^2\}^{2k}$, we have\\
\mbox{}\hfill$\mathrm{Prob}\left(Z_K 
\left(\sum^k_{i=1}a_if_i,\sum^k_{i=1}b_if_i\right)\!=\!
Z_K\left(f_1,\ldots,f_k\right)\right))\!\geq\!\frac{8}{9}$\hfill\mbox{}\\
for any $K\!\in\!\{\C,\C_p\}$.
\end{lemma}
While there are certainly earlier results that are more general than
Lemma \ref{lemma:gh} (see, e.g., \cite[Sec.\ 3.4.1]
{giustiheintz} or \cite[Thm.\ 5.6]{koiran}), Lemma \ref{lemma:gh} is more
direct and self-contained for our purposes. For the convenience of the 
reader, we provide its proof. 

\noindent
{\bf Proof of Lemma \ref{lemma:gh}:} Assume
$f_i(x)\!:=\!\sum^d_{j=0}c_{i,j}x^i$ for all
$i\!\in\!\{1,\ldots,k\}$. Let \linebreak 
$W\!:=\!\left(\bigcup^\ell_{i=1}Z_K(f_i)\right)
\setminus Z_K(f_1,\ldots,f_k)$ and
$\fii(u,\zeta)\!:=\!\sum^k_{i=1}u_if_i(\zeta)$ for any $\zeta\!\in\!W$.
Note that $\#W\!\leq\!kd$ and that for any fixed $\zeta\!\in\!W$,
the polynomial $\fii(u,\zeta)$ is linear in $u$ and not identically zero.
By Schwartz's Lemma \cite{schwartz}, for any fixed $\zeta\!\in\!W$,
there are at most $kN^{k-1}$ points $u\!\in\!\{1,\ldots,N\}^k$ with
$\fii(u,\zeta)\!=\!0$. So then, there at most $dk^2N^{k-1}$ points
$u\!\in\!\{1,\ldots,N\}^k$ with $\fii(u,\zeta)\!=\!0$ for some
$\zeta\!\in\!W$.
                                                                                
Clearly then, the probability that a uniformly randomly chosen pair
$(a,b)\!\in\!\{1,\ldots,N\}^{2k}$
satisfies $\fii(a,\zeta)\!=\!\fii(b,\zeta)\!=\!0$ for some $\zeta\!\in\!W$
is bounded above by $\frac{2dk^2}{N}$. So taking $N\!=\!18dk^2$ we are
done. \qed

\subsection{Review of Riemann Hypotheses} 
\label{sub:riemann} 
Primordial versions of the connection between analysis
and number theory are not hard to derive from scratch and have been known
at least since the 19$\thth$ century. For example, letting
$\zeta(s)\!:=\!\sum^\infty_{n=1} \frac{1}{n^s}$
denote the usual {\bf Riemann zeta function}
(for any real number $s\!>\!1$),
one can easily derive with a bit of calculus (see, e.g., 
\cite[pp.\ 30--32]{des}) that
\[ \zeta(s)\!=\!\prod_{p \text{ prime}}\frac{1}{1-\frac{1}{p^s}}, \text{ and
thus } -\frac{\zeta'(s)}{\zeta(s)}\!=\!\sum^\infty_{n=1}
\frac{\Lambda(n)}{n^s},\]
where $\Lambda$ is the classical Mangoldt function
which sends $n$ to $\log p$ or $0$, according as $n=p^m$
for some prime $p$ (and some positive integer $m$) or not.
For a deeper connection, recall that $\pi(x)$ denotes the
number of primes (in $\N$) $\leq\!x$ and that
the {\bf Prime Number Theorem (PNT)} is
the asymptotic formula $\pi(x)\sim\frac{x}{\log x}$
for \mbox{$x\longrightarrow \infty$.} Remarkably then,
the first proofs of PNT, by Hadamard and de la Vall\'ee-Poussin
(independently, in 1896), were based essentially on the fact that
$\zeta(\beta+i\gamma)$ has {\bf no} zeroes on the vertical line
$\beta\!=\!1$.\footnote{Shikau Ikehara
later showed in 1931 that PNT is in fact {\bf equivalent} to the fact that
$\zeta$ has no zeroes on the vertical line $\beta\!=\!1$ (the proof
is reproduced in \cite{dym}).}
                                                                                
More precisely, writing $\rho\!=\!\beta+i\gamma$ for real $\beta$ and
$\gamma$, recall that $\zeta$ admits an analytic continuation
to the complex plane sans the point $1$
\cite[Sec.\ 2]{des}.\footnote{We'll abuse notation henceforth
by letting $\zeta$ denote the analytic continuation of $\zeta$ to
$\C\setminus\{1\}$.} In particular, the only zeroes of $\zeta$ outside
the {\bf critical strip} $\{\rho\!=\!\beta+i\gamma \; | \;
0\!<\!\beta\!<\!1\}$ are the so-called {\bf trivial} zeroes
$\{-2,-4,-6,\ldots,\}$. Furthermore the zeroes of $\zeta$ in the critical
strip are symmetric about the {\bf critical line} $\beta\!=\!\frac{1}{2}$
and the real axis. The {\bf
Riemann Hypothesis (RH)}, from 1859, is then the following assertion: 
\begin{quote}
{\bf (RH)} All zeroes $\rho\!=\!\beta+i\gamma$ of $\zeta$ with
$\beta\!>\!0$ lie on the critical line $\beta\!=\!\frac{1}{2}$.
\end{quote}
Among a myriad of hitherto unprovably sharp statements in algorithmic
number theory, it is known that RH is true
$\Longleftrightarrow \left|\pi(x)-\int^x_{2} \frac{dt}{\log t}\right|\!=\!
O(\sqrt{x}\log x)$ \cite{des}. In particular, RH is widely agreed to
be the most important problem in modern mathematics. Since May 24, 2000,
RH even enjoys a bounty of one million US dollars thanks to the Clay
Mathematics Foundation.

Let us now consider the extension of RH to primes in arithmetic 
progressions: For any primitive $M\thth$ root of unity $\omega_M$, 
define the {\bf (cyclotomic) Dedekind zeta
function} via the formula $\zeta_{Q(\omega_M)}(s)\!:=\!\sum_{\ca}
\frac{1}{(\cN\ca)^s}$, where $\ca$ ranges over all
nonzero ideals of $\Z[\omega_M]$ (the ring of algebraic integers in 
$\Q(\omega_M)$), $\cN$ denotes the norm function, and $s\!>\!1$
\cite{bs}. Then, like $\zeta$, the function
$\zeta_{\Q(\omega_M)}$ also admits an analytic continuation to 
$\C\!\setminus\!\{1\}$ (which we'll also call $\zeta_{\Q(\omega_M)}$), 
$\zeta_{\Q(\omega_M)}$ has trivial zeroes $\{-2,-4,-6,\ldots,\}$, 
and all other zeroes of $\zeta_{\Q(\omega_M)}$ lie in the critical
strip $(0,1)\times \R$ \cite{lago}. (The zeroes of $\zeta_{\Q(\omega_M)}$ in 
the critical strip are
also symmetric about the critical line $\frac{1}{2}\times\R$ and the real
axis.) We then define the following statement:
\begin{quote}
{\bf (GRH$_{\Q(\omega_M)}$)}\footnote{There is definitely conflicting notation 
in the literature as to what the ``Extended'' Riemann Hypothesis 
or ``Generalized'' Riemann Hypothesis are. We thus hope to dissipate 
any possible confusion via subscripts clearly declaring the field 
we are working with.} 
For any primitive $M\thth$ root of unity $\omega_M$, all the
zeroes $\rho\!=\!\beta+i\gamma$ of $\zeta_{\Q(\omega_M)}$ with $\beta\!>\!0$
lie on the critical line $\beta\!=\!\frac{1}{2}$.
\end{quote}
In particular, letting $\pi(x,M)$ denote the number of primes $p$ 
congruent to $1$ mod $M$ satisfying\\
\scalebox{.95}[1]{$p\!\leq\!x$,
it is known that GRH$_{\Q(\omega_M)}$ is true $\Longleftrightarrow
\left|\pi(x,M)-\frac{1}{\varphi(M)}\int^x_{2} 
\frac{dt}{\log t}\right|\!=\!O\left(\sqrt{x}
(\log x+\log M)\right)$,}\\
where $\varphi(M)$ is the number of $k\!\in\!\{1,\ldots,M-1\}$ 
relatively prime to $M$. (This follows routinely from 
the conditional effective Chebotarev Theorem of \cite[Thm.\ 1.1]{lago}, 
taking $K\!=\!\Q$ and $L\!=\!\Q(\omega_M)$ in the notation there.  
One also needs to recall that the discriminant of $\Q(\omega_M)$ 
is bounded from above by $M^{\varphi(M)}$ \cite[Ch.\ 8, pg.\ 260]{bs}.) 

From the very last estimate, an elementary calculation shows that 
FPT is implied by the truth of the hypotheses $\{GRH_{\Q(\omega_{Q_n})}
\}_{n\in\N}$. However, we point out that FPT can {\bf still} hold even 
in the presence of infinitely many non-trivial zeta zeroes off the critical 
line. For instance, if we instead make the weaker assumption that there is 
an $\eps\!>\!0$ such that all the non-trivial zeroes of 
$\{\zeta_{\Q(\omega_{Q_n})}\}_{n\in\N}$ have real part 
$\leq\!\frac{1}{2}+\eps$, then one can still prove the weaker inequality 
$\left|\pi(x,M)-\frac{1}{\varphi(M)}\int^x_{2}
\frac{dt}{\log t}\right|\!=\!O\left(x^{\frac{1}{2}+\eps}
(\log x+\log M)\right)$ (see, e.g., \cite{strip}). 
Another elementary calculation then shows 
that this looser deviation bound {\bf still} suffices to yield FPT. 
In fact, one can even have non-trivial zeroes of $\zeta_{\Q(\omega_{Q_n})}$ 
approach the line $\{\beta\!=\!1\}$ arbitrarily closely, provided 
they do not approach too quickly as a function of $n$. (See \cite{dzh} 
for further details.)  

\section{The Proofs of Our Main Results}  
\label{sec:proofs1}

\subsection{The Univariate Threshold Over $\Q_p$: Proving Theorem \ref{thm:me}} 
The first assertion rests upon a quantum algorithm for finding the 
multiplicative order of an element of $(\Z/p^\ell\Z)^*$ (see 
\cite{shor,boneh}), once we make a suitable reduction from $\fqp$. 
The second assertion relies on properties of primes in specially chosen 
arithmetic progressions, via our generalization (cf.\ Section 2) 
of an earlier trick of Plaisted \cite{plaisted}. 

\noindent 
\fbox{{\bf Proof of the First Assertion:}}  
First note that it clearly suffices to show that we can 
decide (with error probability $<\!\frac{1}{3}$, say) whether the polynomial 
$f(x)\!:=\!x^d-\alpha$ has a root in $\Q_p$, using a number of qubit operations 
polynomial in $\size(\alpha)+\log d$. (This is because we can divide 
by a suitable constant, and arithmetic over $\Q$ is doable in polynomial 
time.) The case $\alpha\!=\!0$ always results in the root $0$, so let us 
assume $\alpha\!\neq\!0$. Clearly then,  
any $p$-adic root $\zeta$ of $x^d-\alpha$ satisfies 
$d\ord_p\zeta\!=\!\ord_p\alpha$. Since we can compute $\ord_p\alpha$ 
and reductions of integers mod $d$ in $\pp$ \cite[Ch.\ 5]{bs}, we can 
then clearly assume that $d|\ord_p\alpha$ (for otherwise, there 
can be no root over $\Q_p$). Moreover, by rescaling $x$ by an appropriate 
power of $p$, we can assume further that $\ord_p\alpha\!=\!0$. 

Now note that $f'(\zeta)\!=\!d\zeta^{d-1}$ and thus 
$\ord_p f'(\zeta)\!=\!\ord_p(d)$. So 
by Hensel's Lemma, it suffices to decide whether 
the $\mod \ p^\ell$ reduction of $f$ has a root in $\Z/p^\ell\Z$, 
for\linebreak  
$\ell\!=\!1+2\ord_p d$. (Note in particular that $\size(p^\ell)\!=\!
O(\log(p)\log(d))$ which is polynomial in our notion of input size.) By 
Lemma \ref{lemma:fp}, we can easily 
decide the latter feasibility problem, 
given the multiplicative order of $\alpha$ in $(\Z/p^\ell\Z)^*$; 
and we can do the latter in $\bqp$ by 
Shor's seminal algorithm for computing order in a cyclic group 
\cite[pp.\ 1498--1501]{shor}, provided $p^\ell\!\not\in\!\{8,16,32,\ldots\}$. 
So the first assertion is proved for $p^\ell\!\not\in\!\{8,16,32,\ldots\}$. 

To dispose of the remaining cases $p^\ell\!\in\!\{8,16,32,\ldots\}$, 
write $\alpha\!=\!(-1)^a5^b$ and observe that such an expression is 
unique, by the last part of Lemma \ref{lemma:fp}. The first part of 
Lemma \ref{lemma:fp} then easily yields that $x^d-\alpha$ has a root 
iff\\ 
\mbox{}\hfill($a$ odd $\Longrightarrow d$ is odd)$\wedge$(the order of $5^b$ 
divides $\frac{2^{\ell-2}}{\gcd(d,2^{\ell-2})}$).\hfill\mbox{}\\ 
In particular, we see that $x^d-\alpha$ {\bf always} has a root 
when $d$ is odd, so we can assume henceforth that $d$ is even. 

Letting $\flat$ be the order of $5^b$, it is then easy to check that the 
order of $\alpha$ is either $\flat$ or $2\flat$, according as 
$a$ is even or odd. Moreover, since $d$ is even, we see that 
$x^d-\alpha$ can have no roots in $(\Z/2^\ell\Z)^*$ when $a$ is odd. So we 
can now reduce the feasibility of $x^d-\alpha$ to {\bf two} order computations 
as follows: Compute, now via Boneh and Lipton's quantum algorithm for 
order computation in Abelian groups \cite[Thm.\ 2]{boneh}, 
the order of $\alpha$ and $-\alpha$. Observe then that $a$ is odd iff 
the order of $\alpha$ is larger (and then $x^d-\alpha$ has no roots 
in $(\Z/2^\ell\Z)^*$), so we can assume henceforth that $\alpha$  
has the smaller order. To conclude, we then declare that $x^d-\alpha$ has a 
root in $\Q_2$ iff the order of $\alpha$ divides $\frac{2^{\ell-2}}{d}$. This 
last step is correct, thanks to the first part of Lemma \ref{lemma:fp}, 
so we are done.  

\noindent
\fbox{{\bf Proof of the Second Assertion:}} First note that 
$\size(Q_n)\!=\!O(n\log n)$, via the Prime Number Theorem. 
Observe then that the truth of FPH implies
that we can efficiently find a prime $p$ of the form $kQ_n+1$, with
$k\!\in\!\{1,\ldots,2^{n^C}\}$, via random sampling, as follows:
Pick a uniformly randomly integer from $\{1,\ldots,2^{n^C}\}$ 
and using, say, the famous polynomial-time AKS primality testing algorithm
\cite{aks}, verify whether $kQ_n+1$ is prime. We repeat this, no more 
than $9n^{C'}$ times, until we've found a prime. 

Via the elementary estimate $(1-\frac{1}{B})^{Bt}\!<\!\frac{1}{t}$, 
valid for all $B,t\!>\!1$, we then easily obtain that our method 
results in a prime with probability at least $\frac{8}{9}$. 
Since $\size(1+2^{n^C}Q_n)\!=\!O(\log(2^{n^C}Q_n))\!=\!O(n^C+n\log n)$, 
it is clear that our simple algorithm requires a number of bit operations 
just polynomial in $n$. Moreover, the number of random bits needed 
is clearly $O(n^C)$. 

Having now probabilistically generated a prime $p\!=\!1+kQ_n$, 
Lemma \ref{lemma:plai} then immediately yields the implication
``$\fqp(\cU\cS)\!\in\!\cC \Longrightarrow \np\!\in\!\cC\cup\bpp$,'' where
$\cU\cS\!:=\!\{(f_1,\ldots,f_k)\; | \; f_i\!\in\!\Z[x_1] \ , \ k\!\in\!\N\}$:
Indeed, if $\fqp(\cU\cS)\!\in\!\cC$ for some complexity class $\cC$,
then we could combine our hypothetical $\cC$ algorithm for
$\fqp(\cU\cS)$ with our randomized prime generation routine (and the
Plaisted morphism for $K\!=\!\Q_p$) to obtain an  
algorithm with complexity in $\cC\cup \bpp$ for any $\sat$ instance. 
                                                                                
So now we need only show that this hardness persists if we reduce $\cU\cS$ to
systems consisting of just one univariate sparse polynomial. 
Clearly, we can at least reduce to pairs of polynomials 
via Lemma \ref{lemma:gh}, so now we need only reduce from pairs to 
singletons. 

Toward this end, suppose $a\!\in\!\Z$ is a non-square mod $p$ and $p$ 
is odd. 
Clearly then, the only root in $\F_p$ of (the mod $p$ reduction of) the
quadratic form $q(x,y)\!:=\!x^2-ay^2$ is $(0,0)$. Furthermore, by considering
the valuations of $x$ and $y$, it is also easily checked that the only root of
$q$ in $\Q_p$ is $(0,0)$. Thus, given any $(f,g)\!\in\!\Z[x_1]^2$,
we can form $q(f,g)$ (which has size $O(\size(f)+\size(g)+\size(p))$) 
to obtain a polynomial time reduction of 
$\fqp(\Z[x_1]^2)$ to $\fqp(\Z[x_1])$, 
assuming we can find a quadratic non-residue efficiently. 
(If $p\!=\!2$ then we can simply use $q(x,y)\!:=\!x^2+xy+y^2$ 
and then there is no need at all for a quadratic non-residue.) However, this 
can easily be done by picking two random $a\!\in\!\F_p$: 
With probability at least $\frac{3}{4}$, at least one of these 
numbers will be a quadratic non-residue (and this can be checked 
in $\pp$ by computing $a^{(p-1)/2}$ via recursive squaring). 
So we are done. \qed

\subsection{Detecting Square-Freeness: Proving Corollary 1} 
Given any $f\!\in\!\Z[x_1]$, observe that $f$ has a root in 
$\Q_p$ iff $f^2$ is divisible by the square of a degree $1$ polynomial in 
$\Q_p[x_1]$. Moreover, since $\size(f^2)\!=\!O(\size(f)^2)$, we thus 
obtain a polynomial-time reduction of $\fqp(\Z[x_1])$ to the 
problem considered by Corollary 1. So we are done. \qed 

\section*{Acknowledgements} 
The author thanks Leonid Gurvits, Erich Kaltofen, David Alan Plaisted 
for their kind encouragement. Leonid Gurvits 
and Erich Kaltofen also respectively pointed out the references 
\cite{fkw} and \cite{kk}. 

\bibliographystyle{acm}

\end{document}